\documentclass[letterpaper,twocolumn,english,aps,prl,superscriptaddress,floatfix]{revtex4}
\usepackage{amsmath}
\usepackage{graphicx}

\makeatletter
\@ifundefined{textcolor}{}
{%
 \definecolor{BLACK}{gray}{0}
 \definecolor{WHITE}{gray}{1}
 \definecolor{RED}{rgb}{1,0,0}
 \definecolor{GREEN}{rgb}{0,1,0}
 \definecolor{BLUE}{rgb}{0,0,1}
 \definecolor{CYAN}{cmyk}{1,0,0,0}
 \definecolor{MAGENTA}{cmyk}{0,1,0,0}
 \definecolor{YELLOW}{cmyk}{0,0,1,0}
 }

\renewcommand{\[}{\begin{equation}}

\renewcommand{\]}{\end{equation}}

\usepackage{babel}

\usepackage{babel}

\makeatother

\usepackage{babel}

\begin{document}
\global\long\def\avg#1{\langle#1\rangle}

\global\long\def\p{\prime}

\global\long\def\dg{\dagger}

\global\long\def\ket#1{|#1\rangle}

\global\long\def\bra#1{\langle#1|}

\global\long\def\proj#1#2{|#1\rangle\langle#2|}

\global\long\def\inner#1#2{\langle#1|#2\rangle}

\global\long\def\tr{\mathrm{tr}}

\global\long\def\pd#1#2{\frac{\partial#1}{\partial#2}}

\global\long\def\spd#1#2{\frac{\partial^{2} #1}{\partial#2^{2}}}

\global\long\def\der#1#2{\frac{d #1}{d #2}}

\global\long\def\im{\imath}

\global\long\def\As{{^{\sharp}}\hspace{-1mm}\mathcal{A}}

\global\long\def\Fs{{^{\sharp}}\hspace{-0.7mm}\mathcal{F}}

\global\long\def\Es{{^{\sharp}}\hspace{-0.5mm}\mathcal{E}}

\global\long\def\Fd{{^{\sharp}}\hspace{-0.7mm}\mathcal{F}_{\delta}}

\global\long\def\S{\mathcal{S}}

\global\long\def\A{\mathcal{A}}

\global\long\def\F{\mathcal{F}}

\global\long\def\E{\mathcal{E}}

\global\long\def\SgF{\S\left|\F\right.}

\global\long\def\SgEF{\S\left|\E/\F\right.}

\global\long\def\U{\mathcal{U}}

\global\long\def\V{\mathcal{V}}

\global\long\def\H{\mathbf{H}_{\S\E}}

\global\long\def\HSF{\mathbf{H}_{\S\F}}

\global\long\def\HEF{\mathbf{H}_{\S\E/\F}}

\global\long\def\ES{H_{\S}(t)}

\global\long\def\ESo{H_{\S}(0)}

\global\long\def\EgF{H_{\SgF} (t)}

\global\long\def\EgEF{H_{\SgEF} (t)}

\global\long\def\EF{H_{\F}(t)}

\global\long\def\EFo{H_{\F}(0)}

\global\long\def\ESF{H_{\S\F}(t)}

\global\long\def\ESEF{H_{\S\E/\F}(t)}

\global\long\def\ESSEF{H_{\tilde{\S}\S\E/\F}(t)}

\global\long\def\EEFo{H_{\E/\F}(0)}

\global\long\def\EEF{H_{\E/\F}(t)}

\global\long\def\MI{I\left(\S:\F\right)}

\global\long\def\BS{\left\{  \Pi_{j}^{\S}\right\}  }

\global\long\def\QD{\delta\left(\S:\F\right)_{\BS}}

\global\long\def\QDz{\delta\left(\S:\F\right)_{\left\{  \sigma^{z}\right\}  }}

\global\long\def\NQD{\bar{\delta}\left(\S:\F\right)_{\BS}}

\global\long\def\EFS{H_{\F\left| \BS\right. }(t)}

\global\long\def\rhoS{\rho_{\S}(t)}

\global\long\def\rhoSo{\rho_{\S}(0)}

\global\long\def\rhoSF{\rho_{\S\F} (t)}

\global\long\def\rhoSgEF{\rho_{\SgEF} (t)}

\global\long\def\rhoSgF{\rho_{\SgF} (t)}

\global\long\def\rhoF{\rho_{\F}(t)}

\global\long\def\rhoFp{\rho_{\F}(\pi/2)}

\global\long\def\LE{\Lambda_{\E}(t)}

\global\long\def\LEc{\Lambda_{\E}^{\star}(t)}

\global\long\def\LF{\Lambda_{\F}(t)}

\global\long\def\LFc{\Lambda_{\F}^{\star}(t)}

\global\long\def\LEF{\Lambda_{\E/\F} (t)}

\global\long\def\LEFc{\Lambda_{\E/\F}^{\star}(t)}

\global\long\def\Hb{H}

\global\long\def\kE{\kappa_{\E}(t)}

\global\long\def\kEF{\kappa_{\E/\F}(t)}

\global\long\def\kF{\kappa_{\F}(t)}

\global\long\def\ts{t=\pi/2}

\global\long\def\mc#1{\mathcal{#1}}

\renewcommand{\onlinecite}[1]{\cite{#1}}

\title{Quantum Darwinism in a hazy environment}

\author{Michael Zwolak, H. T. Quan, Wojciech H. Zurek}

\affiliation{Theoretical Division, MS-B213, Los Alamos National Laboratory, Los
Alamos, NM 87545}

\date{\today{}}
\begin{abstract}
Quantum Darwinism recognizes that we - the observers - acquire our
information about the ``systems of interest'' indirectly from their
imprints on the environment. Here, we show that information about
a system can be acquired from a mixed-state, or \emph{hazy}, environment,
but the storage capacity of an environment fragment is suppressed
by its initial entropy. In the case of good decoherence, the mutual
information between the system and the fragment is given solely by
the fragment's entropy increase. For fairly mixed environments, this
means a reduction by a factor $1-h$, where $h$ is the \emph{haziness}
of the environment, i.e., the initial entropy of an environment qubit.
Thus, even such hazy environments eventually reveal the state of the
system, although now the intercepted environment fragment must be
larger by $\sim\left(1-h\right)^{-1}$ to gain the same information
about the system.
\end{abstract}
\maketitle
How the classical world arises from an ultimately quantum substrate
has been a question since the advent of quantum mechanics \cite{Schrodinger35-1,Bohr28-1,Einstein35-1,Dirac47-1,Fuchs00-1,vonNeumann32-1,Wheeler84-1}.
Decoherence is now commonly used to study this quantum-classical transition
\cite{Schlosshauer08-1,Zurek03-1,Joos03-1}. Its theory, however,
treats the environment as a sink where information about the system
gets lost forever. Yet the information deposited in the environment
can be intercepted, and it is our primary source of information about
the Universe. Indeed, decohering interactions with the environment
can amplify and store an impression of the system. Amplification was
invoked already by Bohr \cite{Bohr58-1} in the context of measurements.
Early \cite{Zurek82-1}, as well as more recent
\cite{Zurek03-1,Recent,Bennett08-1}, discussions of decoherence
note the importance of redundancy, and provide an information-theoretic
framework for how the environment acts as an amplifier and as a source
of information about the ``system of interest'' \cite{Zurek09-1,Ollivier04-1,Blume05-1,Paz09-1,Brunner08-1}.

Quantum Darwinism reflects this new focus on the environment as a
communication channel \cite{Zurek09-1,Ollivier04-1,Blume05-1}.
When one receives a fragment of the environment by, for instance,
intercepting with one's eyes a portion of photons that are scattered
off a system of interest (e.g., the text of this paper), one acquires
information about it. Previous studies found that, with an initially
pure environment, one can acquire information about the preferred
observables of the system even from small environment fragments \cite{Blume05-1}.
This explains the emergence of objectivity, as it allows many initially
ignorant observers to independently obtain nearly complete information
and reach consensus about the state of the system by intercepting
different fragments of the environment. Classicality of states can
now be quantified in terms of the redundancy of information transferred
to and recorded by the environment. However, it is unclear how well
one can accumulate information starting with a mixed, or \emph{hazy},
environment, such as one at finite temperature. Yet the photon environment
that is responsible for the vast majority of the information we gain
has precisely such a hazy character. This paper shows that even hazy
environments will, in the end, communicate a very clear image.

We study how information flows into a hazy environment using the quantum
mutual information, $\MI=\ES+\EF-\ESF$, between the system, $\S$,
and some fragment, $\F$, of the environment, $\E$. Here, $H\left(t\right)$
is the von Neumann entropy at time $t$ of the subsystem specified
in its subscript. We begin with some general considerations about
information transfer from $\S$ to $\F$ due to a purely decohering
Hamiltonian (i.e., a Hamiltonian, $\H$, that commutes with the preferred
pointer observable of $\S$ \cite{Zurek03-1}). Under evolution generated
by such a Hamiltonian, $\S$ alone, as well as $\S$ plus a small
fragment $\F$ of $\E$, will eventually become effectively decohered
when coupled to $\E/\F$, i.e., the rest of $\E$. In this case of
``good decoherence,'' the state of a qubit $\S$ will evolve as \begin{equation}
\rhoSo=\left(\begin{array}{cc}
s_{00} & s_{01}\\
s_{10} & s_{11}\end{array}\right)\to\rhoS\approx\left(\begin{array}{cc}
s_{00} & 0\\
0 & s_{11}\end{array}\right),\label{eq:decS}\end{equation}
where the final $\rho_{\S}$ is diagonal in its pointer basis (with
obvious generalization to larger system sizes). The system plus the
fragment will therefore become \begin{equation}
\rhoSF\approx\left(\begin{array}{cc}
s_{00}\mathcal{U}_{0}\rho_{\F}(0)\mathcal{U}_{0}^{\dg} & 0\\
0 & s_{11}\mathcal{U}_{1}\rho_{\F}(0)\mathcal{U}_{1}^{\dg}\end{array}\right),\label{eq:RhoSNDecohered}\end{equation}
where $\mathcal{U}_{i}$ is the evolution operator projected onto
the $i^{th}$ pointer state of $\S$. This is because the remaining
portion of the environment, $\E/\F$, suffices to decohere $\S$ plus
$\F$ while preserving the pointer basis of $\S$. The entropy of
the resulting state, Eq. \eqref{eq:RhoSNDecohered}, is thus identical
to the entropy of the state $\rhoS\otimes\rho_{\F}(0)$. The mutual
information between $\S$ and $\F$ then becomes\begin{equation}
\MI=\EF-\EFo\label{eq:GenRel}\end{equation}
in this case of good decoherence. This formula reduces to $\MI=\EF$
for initially pure $\E$ \cite{Zurek07-1}. Equation \eqref{eq:GenRel}
shows that information about $\S$ stored in an initially uncorrelated
fragment $\F$ is represented solely by $\F$'s increase in entropy,
which is due to the interaction with $\S$. Further, it shows how
the capacity for $\F$ to store information is suppressed by its initial
entropy, $\EFo$. In effect, when the initial state of a fragment
$\F$ is completely mixed, it has zero capacity for information as
$\MI=0$ always. Equation \eqref{eq:GenRel} is a general result for
purely decohering Hamiltonians in the case of good decoherence. 

We will now see more explicitly what these results entail by studying
a solvable model of a single spin interacting with an environment
of $\Es$ spins according to the purely decohering Hamiltonian\begin{equation}
\H=\frac{1}{2}\sum_{k=1}^{\Es}\sigma_{\S}^{z}\sigma_{k}^{z}.\label{eq:Hamiltonian}\end{equation}
The advantage of this model is that not only is the evolution solvable
but the entropy of $\F$ can be efficiently computed, enabling the
investigation of large $\E$ (e.g., we have gone up to $\F$ with
$\Fs=200$ spins, giving a Hilbert space dimension of $2^{200}$)
\cite{Quan09-1}. 

We consider here an $\S$ initially uncorrelated with a symmetric
$\E$, i.e., $\rho_{\S\E}(0)=\rhoSo\otimes\rho_{r}^{\otimes\Es}$,
where \[
\rho_{r}=\left(\begin{array}{cc}
r_{00} & r_{01}\\
r_{10} & r_{11}\end{array}\right)\]
 in the $\sigma^{z}$ basis. The haziness, $h$, is the preexisting
entropy of an environment qubit, \[
h=-\tr\left(\rho_{r}\log_{2}\rho_{r}\right).\]
It attains its maximum value of one bit when the qubit is completely
mixed. Other factors, such as the dimension or alignment of the eigenstates
of $\rho_{r}$ with the basis singled out by $\H$, affect the information
acquired by $\E$, as we discuss elsewhere along with our numerical
procedure \cite{Quan09-1}.

We start by elucidating the concept of good decoherence. At time $t$,
the reduced density matrix of $\S$ is \begin{eqnarray}
\rhoS & = & \left(\begin{array}{cc}
s_{00} & s_{01}\LE\\
s_{10}\LEc & s_{11}\end{array}\right),\label{eq:Srho}\end{eqnarray}
where $\LE=\prod_{k\in\E}\Lambda_{k}(t)$ and $\Lambda_{k}(t)=\cos\left(t\right)+\im\left(r_{11}-r_{00}\right)\sin\left(t\right)$
for all $k$ in our model. The decoherence factor $\LE$ represents
how much $\S$ has been decohered by $\E$. The reduced density matrix
$\rhoSF$ has a similar structure, but the decoherence factor is $\LEF$,
i.e., a product of $\Es-\Fs$ instead of $\Es$ individual $\Lambda_{k}(t)$.
For sufficiently small $\Lambda_{k}(t)$ and/or $\Es\gg\Fs$, this
will result in both $\rhoS$ and $\rhoSF$ being nearly diagonal in
the pointer basis of $\S$. This is ``good decoherence.'' Moreover,
$\rhoSF$ can be diagonalized exactly for arbitrary conditions %
\footnote{The unitary $\left(\V\left(t\right)\mathcal{W}\right)^{\otimes\Fs}\oplus\left(\V\left(-t\right)\mathcal{W}\right)^{\otimes\Fs}$
diagonalizes the fragment portion of $\rhoSF$, where $\V\left(t\right)=\exp\left[-\im t\sigma^{z}/2\right]$
and $\mathcal{W}$ diagonalizes $\rho_{r}$. There is a simple generalization
to environments with unequal couplings, etc., which, for pure initial
$\E$, yields the compact expression: $\MI=\Hb\left(\kF\right)+\Hb\left(\kE\right)-\Hb\left(\kEF\right)$.%
}, giving \begin{eqnarray}
\MI & = & \left[\EF-\EFo\right]\nonumber \\
 &  & +\left[\Hb\left(\kE\right)-\Hb\left(\kEF\right)\right],\label{eq:GDGen}\end{eqnarray}
where $\Hb(x)=-x\log_{2}x-(1-x)\log_{2}(1-x)$ and \[
\kappa_{\mathcal{A}}(t)=\frac{1}{2}\left(1+\sqrt{\left(s_{11}-s_{00}\right)^{2}+4\left|s_{01}\right|^{2}\left|\Lambda_{\A}(t)\right|^{2}}\right).\]
In our symmetric model, $\EFo=\Fs\, h$. Equation \eqref{eq:GDGen}
demonstrates that $\MI$ is given exactly by the good decoherence
expression, Eq. \eqref{eq:GenRel}, plus a term giving the deviation
from good decoherence. Essentially everywhere except for short $t$
and $\F\simeq\E$, decoherence is good and the deviation term is nearly
zero since both $\LE$ and $\LEF$ are almost zero. For our model
system with $r_{00}=1/2$, there is a time $t=\pi/2$ when $\Lambda_{k}(t)=0$
and thus the condition for good decoherence is satisfied exactly except
when $\Fs=\Es$.

\begin{figure}[ht]
\noindent \begin{centering}
\includegraphics[width=8cm]{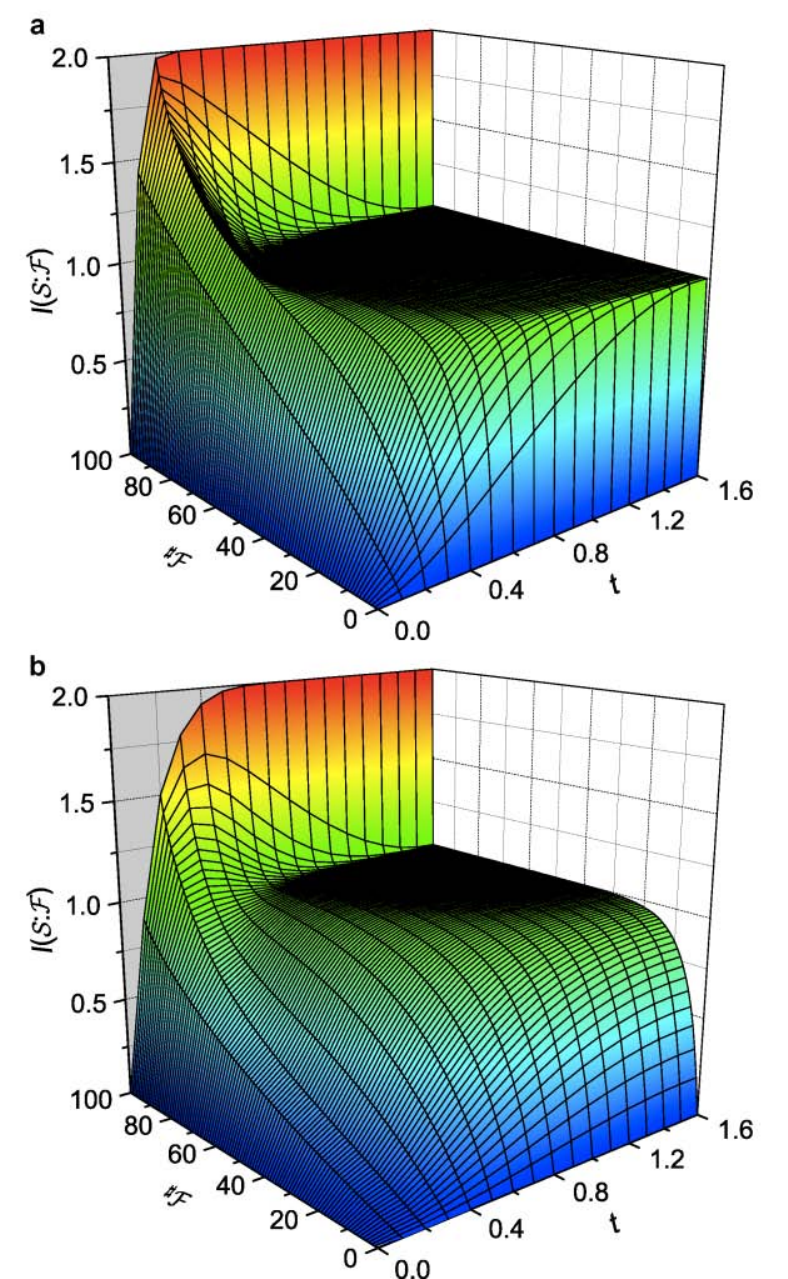}
\par\end{centering}

\noindent \centering{}\caption{Mutual information between $\S$ and a fragment $\F$ for $\Es=100$,
$\rho_{\S}\left(0\right)=\proj ++$, and $r_{00}=1/2$, where $\ket +=\left(\ket 0+\ket 1\right)/\sqrt{2}$.
An initially (a) pure $\E$, $h=0$, and (b) hazy $\E$, $h\approx0.8$.
At $t=0$, $\S$ and $\E$ are not correlated but, with time, correlations
develop. As the system approaches good decoherence, a long plateau
region forms where information is redundantly recorded in $\E$. This
level occurs at the value of the entropy of $\S$ when decohered,
which here is at $H_{\S}=1$. For sufficiently large $\Es$, the mutual
information will eventually reach $H_{S}$ regardless of the initial
haziness, except for a completely mixed initial $\E$, $h=1$. However,
the plateau is attained more slowly and only for larger fragments
as $\E$ gets more mixed. When all of $\E$ is captured, the mutual
information jumps to its maximum value of $2H_{S}$ (signifying complete
quantum correlation of $\E$ with $\S$) so long as a well defined
plateau exists.\label{fig:IvsTime}}

\end{figure}

The evolution of $\MI$ for an initially pure $\E$ is shown in Fig.
\ref{fig:IvsTime}(a). At first, there are no correlations between
$\S$ and $\E$, giving $\MI=0$. As $\S$ decoheres, however, information
is transferred to $\E$. As a consequence, $\MI$ increases. This
increase is initially steep, but the total missing information about
$\S$ is limited by its entropy $H_{\S}$. Therefore, a plateau develops
(the\emph{ }``classical plateau'' \cite{Blume05-1}) as $\MI$ approaches
$H_{S}$. It is seen in Fig. \ref{fig:IvsTime} as the flat region
of the mutual information plot. The level of the plateau occurs at
$H_{\S}=\Hb(s_{00})$. Thus, by intercepting just a few spins from
$\E$ one can gain nearly all the information about $\S$. More precisely,
the redundancy, $R_{\delta}$, is the number of times the state of
$\S$ can be deduced within an accuracy given by the\emph{ information
deficit} $\delta$, i.e., \[
R_{\delta}=\frac{\Es}{\Fd}=\frac{1}{f_{\delta}},\]
where $\Fd$ is the least number of spins needed to acquire a mutual
information greater than $\left(1-\delta\right)H_{S}$ and $f_{\delta}$
is the corresponding fraction of $\E$. For this model, as the time
approaches $\ts$ any single spin from $\E$ has nearly all the information
about $\S$ (i.e., $\delta$ is small for one environment spin). The
redundancy in this case is simply $\Es$, the size of $\E$: many
observers can each capture a single degree of freedom from $\E$ and
gain the same information about $\S$. The existence of the plateau
implies redundancy and demonstrates the validity of the Quantum Darwinism
paradigm. Such a plateau has been found starting with a pure $\E$
in other cases, including models with higher dimensional systems \cite{Ollivier04-1,Blume05-1,Paz09-1}.

The key question we pose here is, what is the effect of starting with
a hazy $\E$? Fig. \ref{fig:IvsTime}(b) plots $\MI$ versus $t$
for $h\approx0.8$. The figure shows that the classical plateau is
still quite large and that it occurs at the same level, $H_{\S}$,
as for an initially pure $\E$. After the plateau is reached, it stays
flat until $\Fs\approx\Es$. The plateau region will always develop
for sufficiently large $\Es$ so long as $\E$ is not totally hazy
($h\neq1$).

\begin{figure}
\noindent \begin{centering}
\includegraphics[width=8cm]{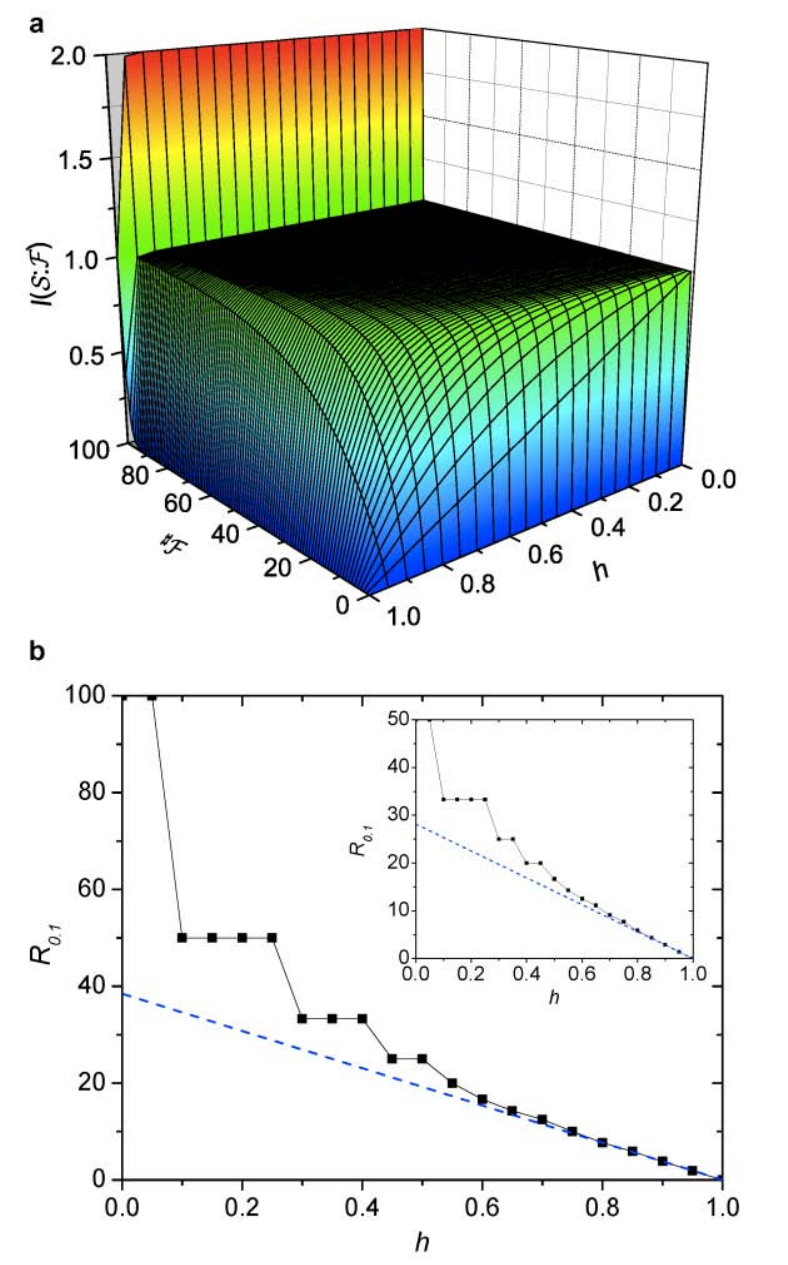}
\par\end{centering}

\noindent \centering{}\caption{(a) Mutual information at $\ts$ versus $\Fs$ and $h$ for the same
conditions as in Fig. \ref{fig:IvsTime}. (b) Redundancy versus $h$
for the information deficit $\delta=0.1$ and at $\ts$ (the inset
shows $t=\pi/3$). The black line (squares) is the exact data. The
redundancy can only take on rational values with $\Es$ in the numerator
because of the discrete nature of the spin environment, which is particularly
visible at high redundancy. The blue dashed line is that obtained
by the scaling $1-h$, which is a good approximation when $h$ is
near one. Thus, even initially mixed $\E$ can store information about
$\S$ in many copies. However, it takes larger $\Fs$ to acquire the
same information about $\S$.\label{fig:Redundancy}}

\end{figure}

These findings are confirmed by an exact result for $\MI$ at $t=\pi/2$
and $r_{00}=1/2$. In this case, $\rho_{\F}\left(\pi/2\right)$ can
be diagonalized exactly %
\footnote{$\F$'s reduced density matrix is given by $s_{00}\otimes_{k\in\F}\left[\V\left(t\right)\rho_{r}\V\left(-t\right)\right]+s_{11}\otimes_{k\in\F}\left[\V\left(-t\right)\rho_{r}\V\left(t\right)\right]$.
At $\ts$ and $r_{00}=1/2$, the two terms are diagonal in the same
basis but with their eigenvalues exchanged.%
}. The entropy of $\F$ is \begin{equation}
H_{\F}(\pi/2)=-\sum_{n=0}^{\Fs}\left(\begin{array}{c}
\Fs\\
n\end{array}\right)\lambda_{\F}(n)\log_{2}\left[\lambda_{\F}(n)\right],\label{eq:ExactSF}\end{equation}
where $n$ labels the degenerate eigenvalues $\lambda_{\F}(n)=s_{00}\lambda_{-}^{n}\lambda_{+}^{\Fs-n}+s_{11}\lambda_{-}^{\Fs-n}\lambda_{+}^{n}$
and $\lambda_{\pm}=1/2\pm\left|r_{01}\right|$ are the eigenvalues
of $\rho_{r}$. Figure \ref{fig:Redundancy}(a) shows $\MI$ versus
$\Fs$ and $h$ at $t=\pi/2$. The plateau region is reached very
rapidly except for $h$ near 1, i.e., very near to a completely mixed
$\E$. The redundancy for the information deficit $\delta=0.1$ is
plotted in Fig. \ref{fig:Redundancy}(b) for $\ts$ and $t=\pi/3$.
For $h$ near 1, the redundancy scales as $R\propto1-h$: The information
storage ability of $\E$ is suppressed by its initial entropy. This
is also reflected in $\MI$ for very mixed states where $\MI\approx\left(1-h\right)\Fs$,
when $\Fs$ is small and under the conditions in the figure. 

\begin{figure}
\begin{centering}
\includegraphics[width=8cm]{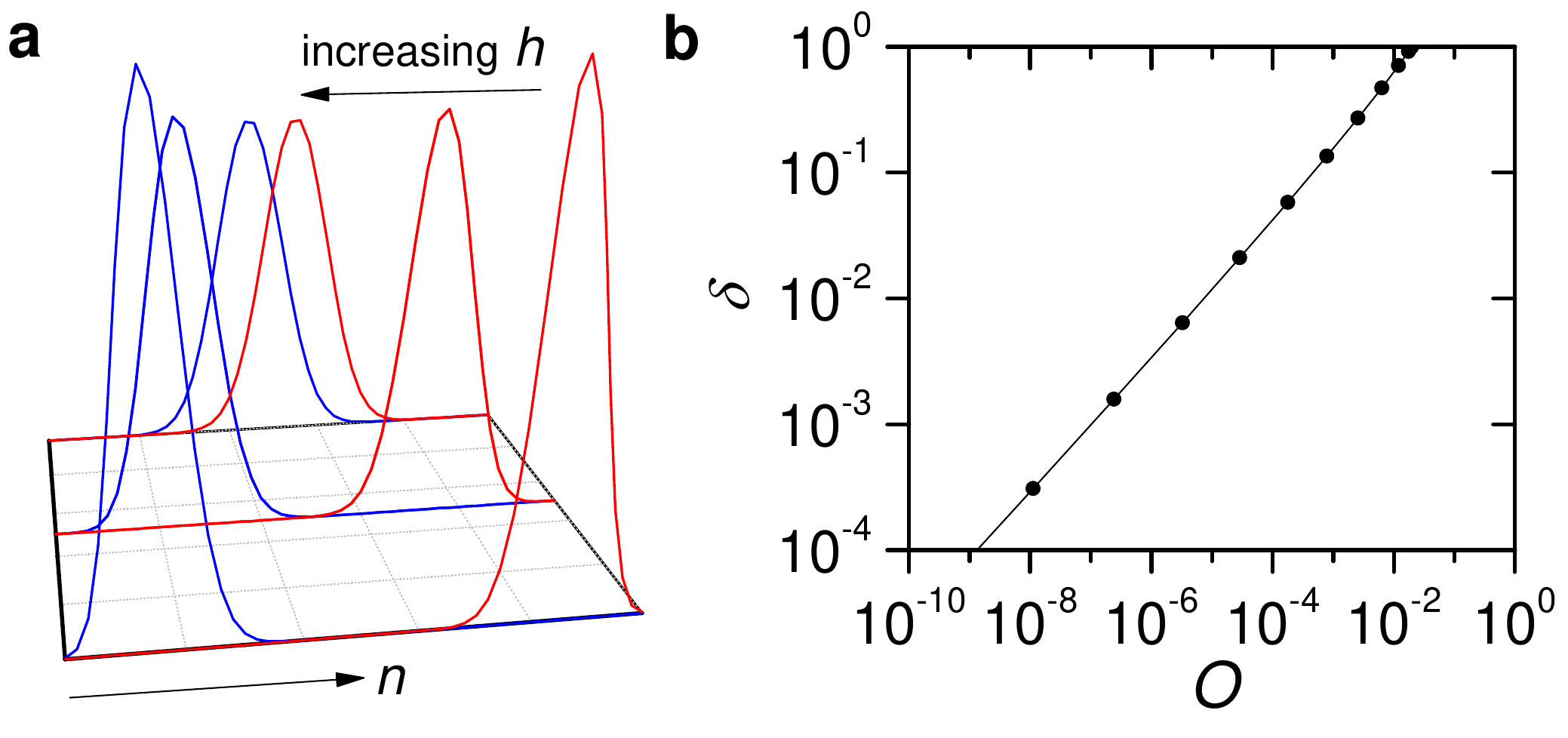}
\par\end{centering}

\noindent \centering{}\caption{(a) Bimodal probability distribution for the state of $\F$ to be
in a subspace $n$, where $n=0,\ldots,\Fs$ indexes the number of
identical records. The left (blue) peak, $P_{L}$, is correlated with
$\ket 0$ in $\S$ and the right (red), $P_{R}$, with $\ket 1$.
When $\E$ is initially nearly pure, $h\sim0$, the two peaks are
well separated. Upon increasing $h$, the two peaks start to overlap.
(b) The information deficit $\delta$ versus the overlap $O=\sum_{n}P_{L}(n)P_{R}(n)$
for $\Fs=50$ and different $h$. A larger overlap is associated with
a decreased capacity for information about $\S$, i.e., an increased
information deficit. The parameters are the same as Fig. \ref{fig:IvsTime}.
\label{fig:bimodal}}

\end{figure}

These results indicate that a hazy $\E$ can be thought of as a noisy
communication channel with degraded capacity, $1-h$. The loss of
the channel capacity can be depicted in terms of the overlap of two
peaks in a bimodal probability distribution over subspaces of $\F$
with $n$ identical records indicating a $\sigma^{y}$ eigenstate
and $\Fs-n$ records indicating the orthogonal eigenstate. At $t=\pi/2$
the probability distribution is given by the two peaks \begin{equation}
P_{L}(n)=s_{00}\left(\begin{array}{c}
\Fs\\
n\end{array}\right)\lambda_{-}^{n}\lambda_{+}^{\Fs-n}\tag{12a}\end{equation}
 and \begin{equation}
P_{R}(n)=s_{11}\left(\begin{array}{c}
\Fs\\
n\end{array}\right)\lambda_{-}^{\Fs-n}\lambda_{+}^{n},\tag{12b}\end{equation}
with $n=0,\ldots,\Fs$. The bimodal structure is the result of information
about $\S$ branching into sectors of $\E$'s Hilbert space. For initially
very pure $\E$ and reasonably large $\Fs$, these two sectors are
distinct, allowing one to resolve the peaks, as shown in Fig. \ref{fig:bimodal}(a).
In this case, $\MI$ is near its plateau value. As the peak overlap
increases, the evidence about the state of $\S$ imprinted in $\F$
becomes less conclusive. Thus, a more hazy environment (Fig. \ref{fig:bimodal}(b))
or a smaller fragment result in an increased information deficit.

To conclude, we studied Quantum Darwinism in the case of a hazy environment.
For good decoherence and purely decohering Hamiltonians, we demonstrated
that the mutual information acquired by some fragment of the environment
is directly related to the entropy increase of that fragment. This
shows that the capacity of the environment to accept information is
suppressed by its initial entropy. Thus, a hazy environment acts like
a noisy communication channel, transmitting all the information about
the system, but at a lower rate. By examining a model system, we illustrated
that, despite this diminished channel capacity, the region of redundant
information storage is still reached for quite mixed environments.
This work leads to questions related to recent research on representing
environments in compact forms in order to accurately simulate a dynamical
quantum system \cite{Zwolak08-1}: What role does the information
acquired play in the representation of the environment, and, vice
versa, does a compact representation yield the same redundancy as
the full $\E$? Above all, however, our results verify the environment's
capability to communicate information.
\begin{acknowledgments}
We would like to thank G. Smith, J. Yard, and M. Zubelewicz. This
research is supported by the U.S. Department of Energy through the
LANL/LDRD Program.
\end{acknowledgments}

\end{document}